\documentclass[aps,prx,twocolumn,superscriptaddress,groupedaddress]{revtex4}  
\usepackage{graphicx}  
\usepackage{dcolumn}   
\usepackage{bm}        
\usepackage{amssymb}   
\usepackage{amsmath}   
\usepackage[mathscr]{euscript}
\usepackage{color}
\usepackage{ulem}
\begin{document}

\title{Nonequilibrium theory of enzyme chemotaxis and enhanced diffusion}
\author{D. M. Busiello}
\affiliation{Institute of Physics, Ecole Polytechnique F\'ed\'erale de Lausanne (EPFL), 1015 Lausanne, Switzerland}
\author{P. De Los Rios}
\affiliation{Institute of Physics, Ecole Polytechnique F\'ed\'erale de Lausanne (EPFL), 1015 Lausanne, Switzerland}
\affiliation{Institute of Bioengineering, Ecole Polytechnique F\'ed\'erale de Lausanne (EPFL), 1015 Lausanne, Switzerland}
\author{F. Piazza}
\affiliation{Centre de Biophysique Mol\'eculaire (CBM), CNRS-UPR 4301, Rue C. Sadron, Orl\'eans 45071, France}
\affiliation{Universit\'e d'Orl\'eans, Ch\^ateau de la Source, Orl\'eans Cedex 45071, France}

\date{\today}

\begin{abstract}
Enhanced diffusion and anti-chemotaxis of enzymes have been reported in several experiments in the last decade, opening up entirely new avenues of research in the bio-nanosciences both at the applied and fundamental level. Here, we introduce a novel theoretical framework, rooted in non-equilibrium effects characteristic of catalytic cycles, that explains all observations made so far in this field. In addition, our theory predicts entirely novel effects, such as dissipation-induced switch between anti-chemotactic and chemotactic behavior.
\end{abstract}

\maketitle

The possibility of designing efficient nano or micro-motors based on enzymes' kinetics has been intensely investigated in recent years \cite{1gole, sen, guha}. Enhanced diffusion and directed motion are essential features hinting at the idea of autonomous machines that would find applications in pattern formation, transport, and sensing problems \cite{6sen,7sen,8sen}. In several contexts, the achievement of directed motion is associated with a catalytic transformation of chemical energy into mechanical force \cite{rsen1,rsen2}. However, it is still unclear under which conditions an enzyme tends to move towards (chemotaxis) or away from (anti-chemotaxis) the substrate, and claims of both these conflicting evidences have been reported in the literature \cite{sen, granick, guha, zhang}. Furthermore, the intimate connection between enhanced diffusion and enzyme taxis is a subject of discussion, although it is widely acknowledged that non-equilibrium features of the catalytic reaction play a fundamental role \cite{granick,goleNE}.

In this Letter, we aim at explaining the results of the experiments reported in \cite{granick}, in which the kinetics of two different catalytic enzymes is studied, showing an anti-chemotactic behavior in a substrate gradient. At the same time, diffusion is enhanced proportionally to the substrate concentration at any point in space. Another interesting feature reported in \cite{granick} is the signature of a ballistic-to-diffusive transition at short times.

Recent work \cite{golestanian} addressed the same problem by constructing a detailed model comprising non-specific interactions and complex formation between enzyme and substrate. These two different interaction mechanisms were found to lead to two competing contributions to the diffusion enhancement. Despite the great accuracy of the microscopic description, this model rests on the hypothesis that the free enzyme and the complexes should have markedly different diffusion coefficients, an unlikely situation in most of the experimental settings studied here \cite{granick,sen}. In particular, in the experiments reported in \cite{granick}, both the enzymes analyzed are much bigger than their substrate. Hence, the diffusion coefficient can be considered independent of the chemical state of the system (free enzyme, enzyme with substrate or enzyme with product). Moreover, in \cite{golestanian}, neither the stationary profiles of both diffusion coefficient and enzyme concentration are shown to be correctly predicted, nor a ballistic-diffusive transition at short times is discussed.

\begin{figure}[t]
\includegraphics[width = 1 \columnwidth]{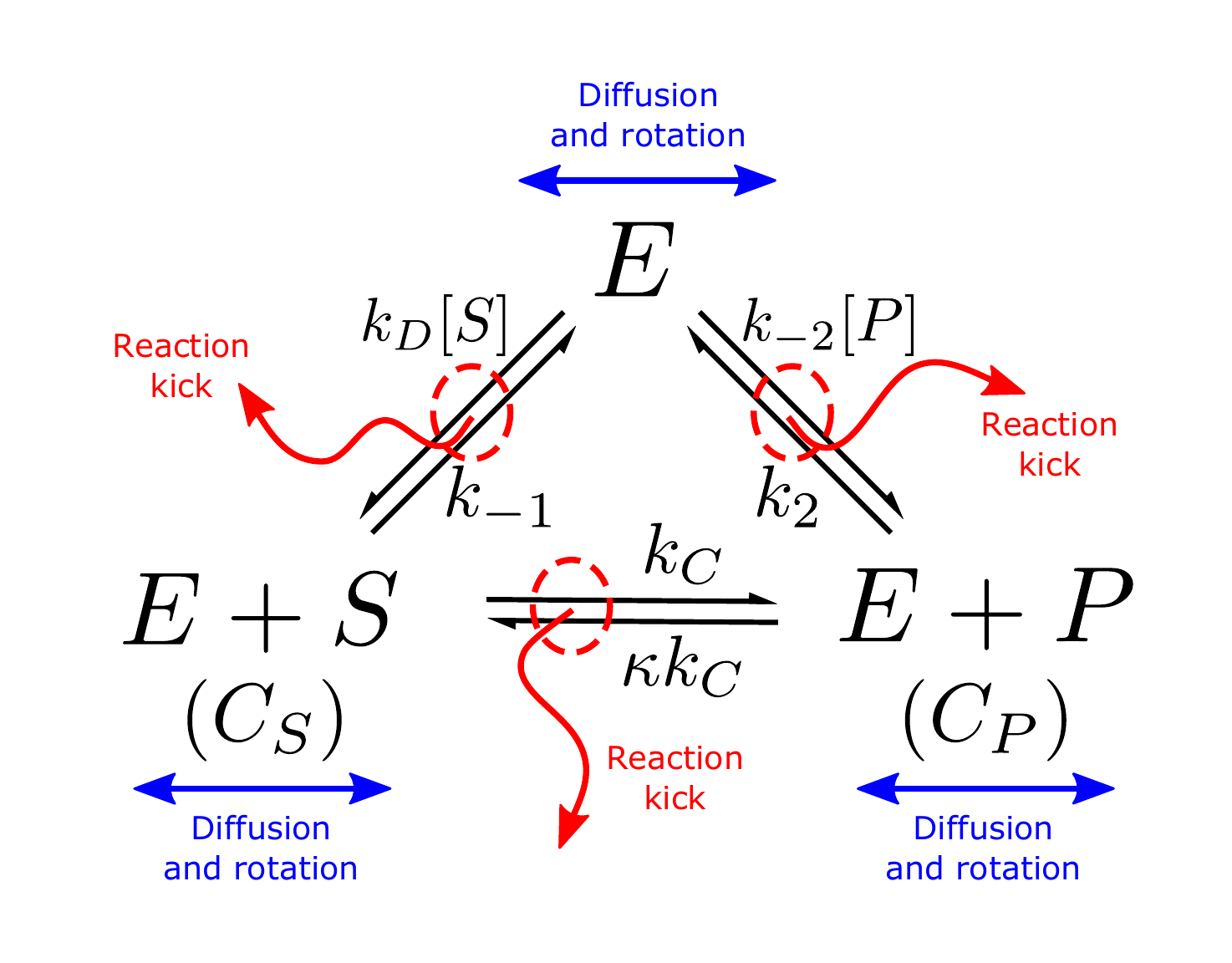}
\caption{3-state chemical system describing a catalytic cycle with diffusion. $E$ is the free enzyme, $C_S$ the complex with the substrate and $C_P$ the complex with the product. Each state has its own orientation (right or left) that can change in time because of a \textit{rotation} mechanism. Heat released during catalysis \citep{bustamante}, charge displacement \cite{ross} or conformation changes \cite{golestanian} can induce \textit{reaction} kicks.}
\label{fig:model}
\end{figure}

Here, we describe the enzyme-substrate-product system as a 3-state chemical reaction model with catalysis. Each state diffuses in a one-dimensional space with the same diffusion rate $d$. Moreover, we include the presence of kicks alongside each chemical transformation. These can be due to the heat released during the catalytic step, as proposed in \cite{bustamante}, to impulsive forces arising from transient electric charge unbalance around the catalytic site \cite{ross}, or even to small conformational changes \citep{golestanian}, whose effect on the diffusion time-scale can be neglected. Here, we do not exclude or endorse any particular mechanism, generically referring to them as \textit{reaction} kicks. While additional experiments appear necessary to elucidate the physical origin of kicks, we show that our model is robust under different choices for what concerns the explanation of the data presented in \cite{granick}.

The anisotropy of the enzyme binding site introduces a natural orientation for each state. In a one-dimensional domain, it can be \textit{towards the right}, $R$, or \textit{towards the left}, $L$. Each orientation is associated with a \textit{preferential} direction of motion, as explained in \cite{goleP1,michelin1,michelin2}. In the model, this effect is encoded in an effective parameter $q$, which is the rate of changing orientation, and hence preferential direction of motion, equal for all the states for simplicity (see Fig. \ref{fig:model}). The simplest microscopic description is given in terms of the probability to be in a given chemical state $X$, going towards a given direction ($R$ or $L$), for each position $x$ and time $t$: $P_{R,L}^{(X)}(x,t)$. We denote by $E$ the free enzyme, $C_S$ the complex with the substrate, and $C_P$ the complex with the product. As an example, we report below the Master Equation for $P_R^{(E)}$ (see SI):
\begin{gather}
P_R^{(E)}(x,t + \Delta t) = P_R^{(E)}(x,t) \Big( 1 - k_{\rm out} \Delta t \Big) + \nonumber \\
+ (1-q) \Delta t \Big( k_{-1} P_R^{(C_S)}\left( x - l_S \Delta x \right) + k_{2} P_R^{(C_P)}\left( x - l_P \Delta x \right) \Big) + \nonumber \\
+ q \Delta t \Big( k_{-1} P_L^{(C_S)}\left( x - l_S \Delta x \right) + k_{2} P_L^{(C_P)}\left( x - l_P \Delta x \right) \Big) + \nonumber \\
+ d \Delta t \left( (1-q) P_R^{(E)}(x - \Delta x) + q P_L^{(E)}(x - \Delta x) \right)
\label{ME}
\end{gather}
with:
\begin{gather*}
k_{\rm out} = k_D [S(x)] + k_{-2} [P(x)] - 2 d \nonumber \\
\end{gather*}
Here, $l_S$ and $l_P$ quantify the extent of kicks in units of $\Delta x$. We remark that all the terms contributing to $P_R^{(E)}(x, t+\Delta t)$ come from the left, i.e. from $x' < x$. The parameter $q$ quantifies the probability that the system switched preferential direction of motion, by changing orientation: it reached $x' < x$ at time $t$ from the right, and then $x$ at time $t + \Delta t$ from the left. Analogous equations can be written for all other states. The mathematical paradigm underlying Eq.~\eqref{ME} is known as persistent random walk \cite{per1, per2}. The word \textit{persistent} is reminiscent of the fact that, depending on the orientation, the system tends to move in a preferred direction.

We perform the continuum limit on the Master Equation, by letting $\Delta t \to 0$, $\Delta x \to 0$ and $q \to 0$, with the following constraints:
\begin{gather}
\lim_{\Delta x ,\Delta t\to 0} d \Delta x = v_d \qquad
\lim_{q,\Delta t \to 0} q d = \tau_r^{-1} \nonumber \\
\lim_{\Delta x,\Delta t \to 0} k_{XY} l_{XY} \Delta x = k_{XY} \lambda_{XY} \;\;\; \forall \; X,Y \nonumber \\
\lim_{q, \Delta t \to 0} q k_{XY} = k_{XY} \nu_{XY} \;\;\; \forall \; X,Y
\label{constraints}
\end{gather}
where $v_d$ is the velocity associated with diffusion, $\tau_r^{-1}$ is the rotation rate, $\lambda_{XY}$ and $\nu_{XY}$, respectively, quantify the size of kicks and the interplay between reaction and rotation during the transformation $X \to Y$. Additionally, $\lambda_{XY} = \lambda_{YX}$ due to the scallop theorem \cite{purcell}. Notice that both chemical ($k_{XY}$, $\forall X,Y$) and diffusive ($d$) rates should be intended as parameters scaling as $(\Delta t)^{-1}$.


Because of kicks, each chemical reaction has a local and a non-local term in the continuum limit. Since Eqs. \eqref{constraints} set the latter to be finite, the local chemistry must take place on a much faster time-scale in this limit. To be consistent, we perform a time-scale separation through the following \textit{ansatz}:
\begin{equation}
\label{fastreac}
P_{R,L}^{(X)}(x,t) = \pi^{(X)}(x) p_{R,L}(x,t)
\end{equation}
where $\pi^{(X)}(x)$ is the stationary solution of the chemical system without kicks, i.e. locally.

Performing the continuum limit, and exploiting Eq. \eqref{fastreac}, we find the following set of equations (see SI):
\begin{eqnarray}
\left( \partial_t - \partial_x \langle L \rangle \right) \frac{1}{\langle N \rangle} \left( \partial_t + \partial_x \langle L \rangle \right) p_R + 2 \partial_t p_R &=& 0 \nonumber \\
\left( \partial_t + \partial_x \langle L \rangle \right) \frac{1}{\langle N \rangle} \left( \partial_t - \partial_x \langle L \rangle \right) p_L + 2 \partial_t p_L &=& 0
\label{teq}
\end{eqnarray}
where $\langle \cdot \rangle$ indicates the average over steady-state local distributions of chemical states, i.e. 
\begin{gather*}
\langle L \rangle = \sum_{XY} k_{XY} \lambda_{XY} \pi^{(X)} + v_d \nonumber \\
\langle N \rangle = \sum_{XY} k_{XY} \nu_{XY} \pi^{(X)} + \tau_r^{-1}
\end{gather*}
We are interested in the probability of finding the enzyme at position $x$ at stationarity, independently of the chemical state and the orientation. This is equal to $P^{ss}(x) = p_R(x,t\to \infty) + p_L(x,t\to \infty)$, and satisfies:
\begin{equation}
\partial_x \left( \frac{\langle L \rangle}{\langle N \rangle} \partial_x \Big( \langle L \rangle P^{ss} \Big) \right) = 0 \quad \to \quad P^{ss}(x) \propto \langle L \rangle^{-1}
\label{sseq}
\end{equation}
In the SI we detail the derivation of Eq. \eqref{sseq}.

To obtain the long-time diffusive behavior of the system, we first perform the diffusive limit on Eqs.~\eqref{teq}, which corresponds to $\langle N \rangle \gg t^{-1}$. The correct procedure is explained in details in the SI. Then, the effective space-dependent diffusion coefficient is estimated, for each point in space, $x^*$, considering substrate and product concentrations to be fixed at their value in $x^*$, $[S(x^*)]$ and $[P(x^*)]$ respectively. We name this quantity $D^{\rm eff}(x^*)$. From Eq. \eqref{sseq}, it follows
\begin{equation}
D^{\rm eff}(x) = \langle L \rangle^2 \left( 2 \langle N \rangle \right)^{-1}
\label{deff}
\end{equation}
where $\langle L \rangle$ and $\langle N \rangle$ depend on space through substrate and product concentrations, $[S(x)]$ and $[P(x)]$. In the absence of substrate, $D^{\rm eff} = v_d^2 ~\tau_r / 2 = D_0$.

In order to compare the theoretical predictions for $P^{ss}(x)$ and $D^{\rm eff}(x)$ with experimental data, we need to identify which parameters can be estimated \textit{a-priori} and which cannot. Consequently, the latter are fitted constraining their values within reasonable intervals.

Consider the following assumption:
\begin{equation}
\frac{[S(x)]}{[P(x)]} = e^{\Delta S_m} \frac{[S]^{\rm eq}}{[P]^{\rm eq}} = e^{\Delta S_m} \mathcal{R}^{\rm eq}
\label{energy}
\end{equation}
where $\Delta S_m > 0$ is the non-dimensional entropy change in the environment during one catalytic cycle, which can be interpreted as the amount of energy maintaining the system out of equilibrium in units of $k_B T$. Since the local chemical reactions are very fast, Eq. \eqref{fastreac}, we assume that the substrate-to-product ratio does not depend on space. The gradient $[S(x)]$ is fixed by the experiment, while $\Delta S_m$ has to be inferred from the data.

For the sake of simplicity, and aiming at reducing the number of free parameters, we set $\lambda_{XY} = \lambda$ and $\nu_{XY} = \nu$, $\forall X,Y$. Moreover, some reasonable $\textit{a-priori}$ values can be fixed:
\begin{equation}
k_2 \approx 10^4 s^{-1} \;\;\;\;\; k_D \approx 10^7 s^{-1} M^{.1} \;\;\;\;\; k_{-1} \approx 10^2 s^{-1}
\end{equation}

As detailed in the SI, the model here presented depends on the chemical rates, on the diffusion coefficient without substrate $D_0$, on the entropy change $\Delta S_m$, and on the dimensionless parameters $\alpha_S$ and $r$, defined as:
\begin{gather}
\alpha_S = \frac{\lambda}{\lambda_0} \;\;\;\;\; \lambda_0 \approx \frac{v_d}{2 k_2} \;\;\;\;\; r = \frac{\nu}{\lambda} v_d \tau_r
\label{pars}
\end{gather}
Here, $\lambda_0$ is a typical length-scale associated with the interplay between the two non-local mechanisms: diffusion and reaction kicks. Assuming that these are of the same order of magnitude, we further constrain $\alpha_S$ to be approximately unity. This is consistent with the interpretation of $\alpha_S$ as a version of the Damk$\ddot{\rm o}$hler number \cite{1974,goppel} in the presence of kicks.

\begin{figure}[t]
\includegraphics[width=\columnwidth]{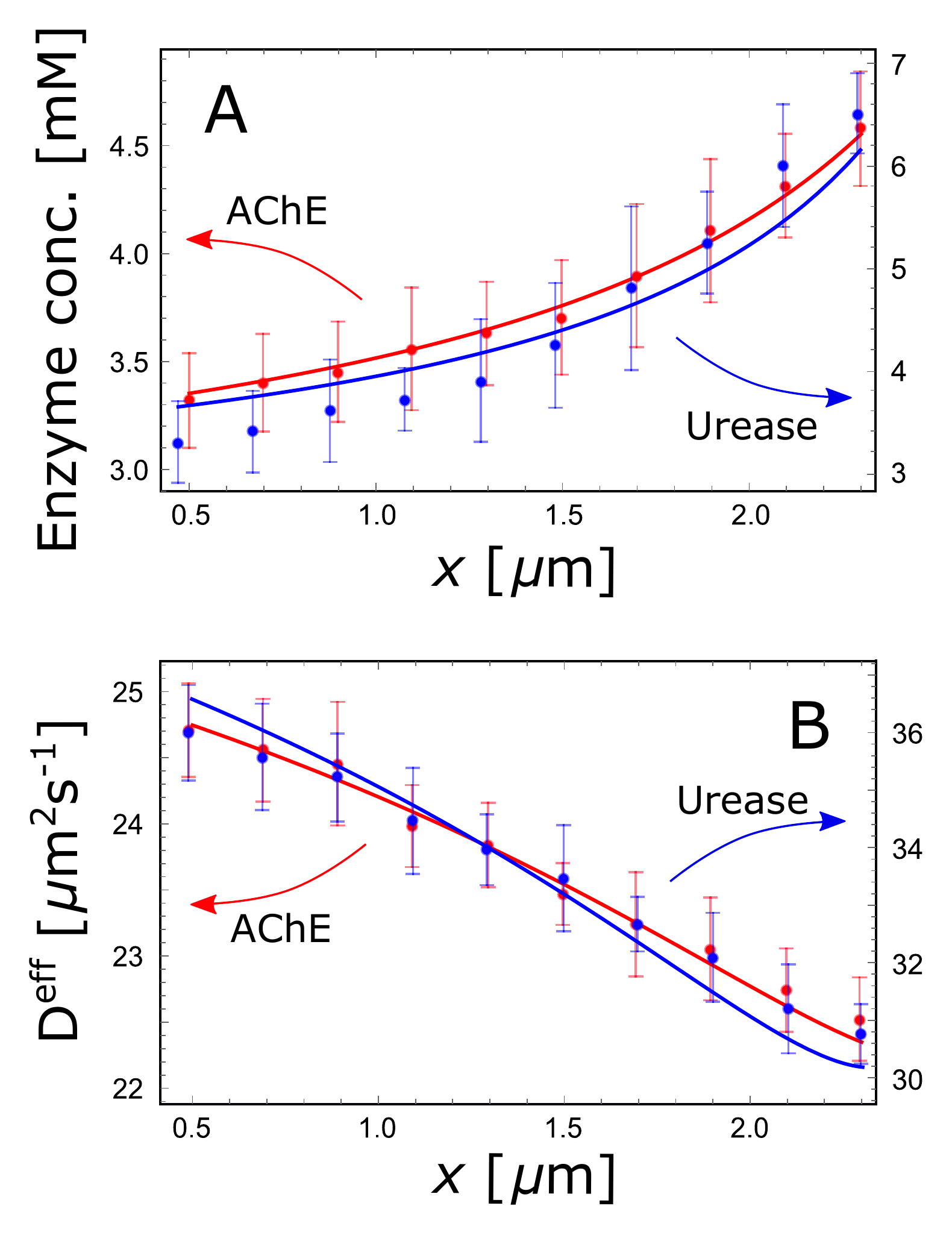}
\caption{\textit{Panel A} - Comparison between data extracted from \cite{granick} (dots) and theoretical predictions, Eq.~\eqref{sseq} (line) for the stationary concentration profile of acetylcholinesterase (AChE) (in red) and urease (in blue). Vertical bars indicate the experimental error. \textit{Panel B} - Comparison between data (dots) and theory, Eq.~\eqref{deff} (line) for the profile of effective diffusion coefficient of AChE (in red) and urease (in blue).}
\label{fig1}
\end{figure}

In Fig. \ref{fig1} we represent the profiles of both concentration and effective diffusion coefficient predicted by the model in comparison with experimental data reported in \cite{granick}, for acetylcholinesterase (AChE) catalyzing acetylcholine hydrolysis (in red) and urease catalyzing urea hydrolysis (in blue). The agreement is striking for AChE and good, within the experimental errors, for urease. The values of all fitted parameters are compatible with physical constraints, as shown in Table \ref{table1}. In the SI we show that fits of comparable quality can be obtained assuming that there are only heat-induced kicks \citep{bustamante} during hydrolysis and synthesis of the substrate. Measurements of stationary profiles performed on non-catalytic molecules could shed further light on the leading mechanism responsible for the presence of kicks.


It is important to remark that, without the catalytic step, i.e. $C_S \leftrightarrow C_P$, it is not possible to find suitable values for the model parameters to have a satisfying agreement between data and theory. Hence, the out-of-equilibrium nature of the system is unveiled by the necessity of dissipating energy along the catalytic cycle, and translates into a non-zero entropy change in the environment, $\Delta S_m$ (see Table \ref{table1}). However, molecular anti-chemotaxis does need to feed upon a substrate concentration gradient, but it does not require the presence of catalytic cycles. In the SI, we show that binding/unbinding chemical systems still exhibit anti-chemotactic profiles and enhanced diffusion under similar working conditions, albeit with substantial differences with respect to the data reported in \cite{granick} and displayed here in Fig. \ref{fig1}.

\begin{table}[t]
\includegraphics[width=\columnwidth]{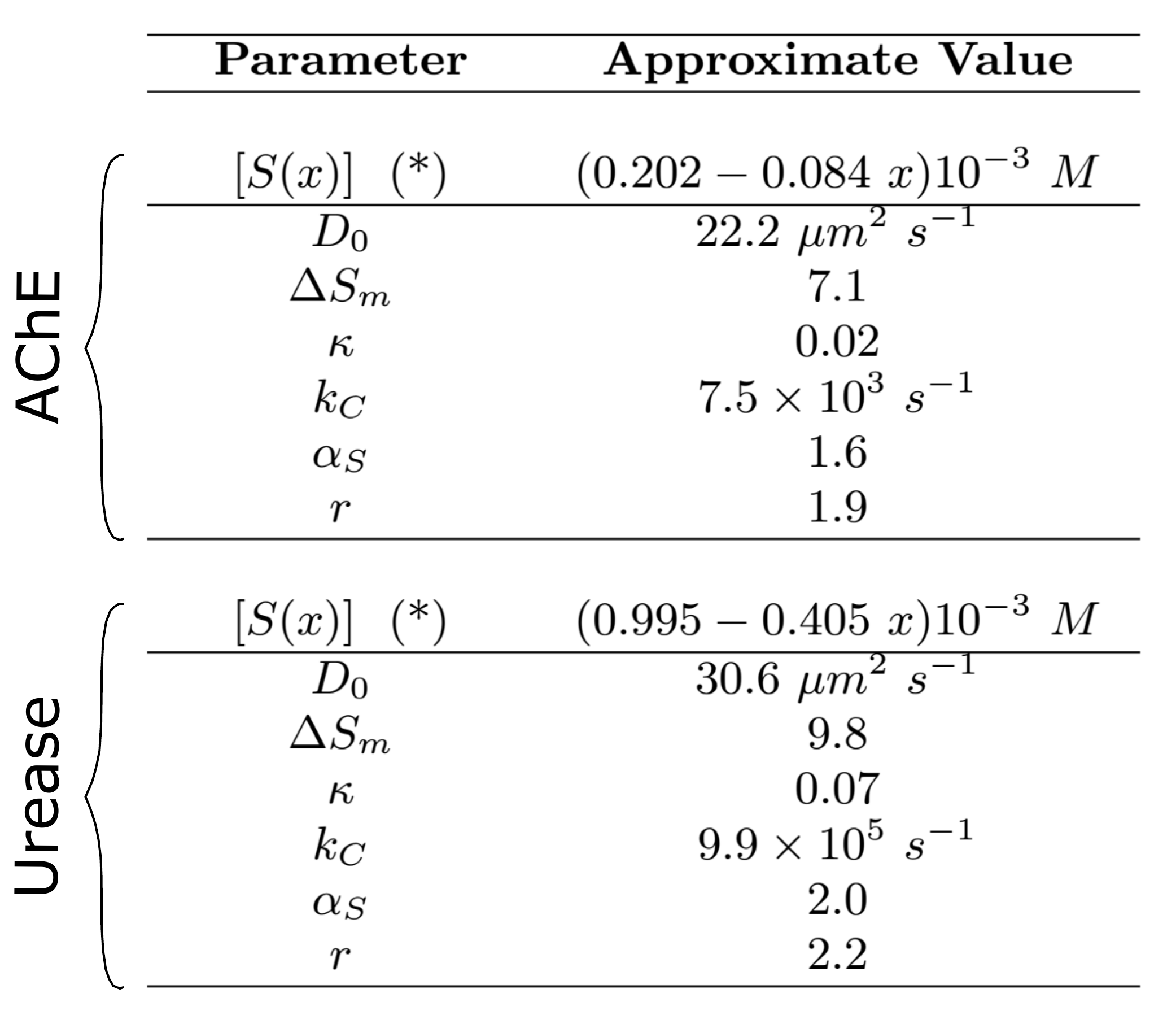}
\caption{List of parameters with their approximate value. The symbol (*) indicates parameters directly obtained from the data, without employing a fitting procedure. The feasibility of the fitted chemical rates is supported by values from the literature \cite{granick, chemrev}. Model-dependent parameters, $\alpha_S$ and $r$, coincide with the physical expectation detailed in the text. The bare diffusion coefficient $D_0$ lies in the range of measured values, according to \cite{granick}.}
\label{table1}
\end{table}

Within our model, a ballistic-to-diffusive transition emerges naturally and can be discussed in the simplified case of a constant substrate concentration, $[S]$. In this scenario, the dynamical evolution of $P(x,t) = p_R(x,t) + p_L(x,t)$ simplifies to
\begin{equation}
\left( 2 \langle N \rangle \right)^{-1} \partial_{tt} P + \partial_t P - D^{\rm eff} \partial_{xx} P = 0
\label{teleg}
\end{equation}
Eq.~\eqref{teleg} is known in the literature as telegrapher's equation, and it exhibits a ballistic-to-diffusive transition for $t = t^* \approx \left( 2 \langle N \rangle \right)^{-1}$ \cite{per1}. Notice that $t^*$ depends on a free parameter $\tau_r$, which can be determined according to the experimental value of the transition time in the constant substrate case. Remarkably, in our model the presence of a ballistic regime at short times appears even in the absence of a substrate gradient. Further experiments are needed to confirm this theoretical prediction. In the SI, we show that a ballistic regime at short times is found also when the system is placed in a gradient of substrate concentration.

We now proceed to show how our model can be employed to resolve a seemingly contradictory observation reported in \cite{sen}. The authors claim that the enzyme (catalase or urease) spreads towards regions of higher substrate concentration, showing molecular chemotaxis. In the experimental setting, a longitudinal laminar flow is sustained by the continuous injection of substrate and enzyme, which mix by diffusion along the transversal direction. Starting from the following initial concentration profiles:
\begin{gather*}
[S(x,0)] = \begin{cases} [S_0], & \mbox{if } 0 \leq x \leq L/2 \\ 0, & \mbox{if } L/2 < x \leq L \end{cases} \\
[E(x,0)] = \begin{cases} 0, & \mbox{if } 0 \leq x \leq L/2 \\ [E_0], & \mbox{if } L/2 < x \leq L \end{cases} \\
[C_S(x,0)] = 0 \;\;\;\;\;\;\; [C_P(x,0)] = 0
\end{gather*}
where $L$ is the transversal dimension of the capillary, the enzyme is observed to move towards the region where the substrate is more abundant.

We consider the same initial conditions. The substrate diffuses much faster than the enzyme, eventually being uniformly distributed along the whole transverse capillary length. At first glance, and in the spirit of a time-scale separation approach, we consider the dynamics of the enzyme, after a transient time, as if it were in the presence of a uniform substrate concentration. The evolution is dominated by a diffusive behavior. The profile obtained in this way, shown in the SI, is qualitatively very similar to the one measured in \cite{sen}. Hence, the chemotactic behavior at the initial stage of the dynamics is the logical transient evolution of the initial conditions \textit{en route} to a flat profile with enhanced diffusion, rather than an intrinsic property of the system. It is captured by our model, which leads to molecular anti-chemotaxis in the long-time limit. 

Our theory can be employed to investigate the conditions under which an enzyme preferentially moves towards or away from the substrate. To this aim, we inspect the possibility to observe a chemotactic stationary profile within the proposed framework. Consider the presence of a gradient of substrate concentration that decreases along the one-dimensional domain: $\partial_x [S(x)] < 0$, $\forall x$. Molecular chemotaxis is characterized by a steady-state probability distribution of enzymes $P^{\rm ch}(x)$ with the same monotonicity as $[S(x)]$: $\partial_x P^{\rm ch}(x) < 0$, $\forall x$. The latter condition cannot be satisfied in the working conditions described so far, for any choice of the free parameters.

In \cite{golestanian}, the difference between complexes and free-enzyme \textit{diffusiophoretic} drift leads to an extra contribution both to enhanced diffusion and to the stationary behavior of the system. Inspired by this observation, we revert to a slightly more general version of our model where the diffusion rate is different depending on the chemical species. We define:
\begin{gather*}
\lim_{\Delta x, \Delta t \to 0} d^{(X)} \Delta x = v_d^{(X)} \qquad
\lim_{q, \Delta t \to 0} q d^{(X)} = 1/\tau_r^{(X)}
\end{gather*}
with $X = E, C_S, C_P$. For simplicity, we further impose $v_d^{(C_S)} = v_d^{(C_P)} \equiv v_d^{(C)}$, and $r^{(X)} \equiv r$, $\forall X$ (see Eq.~\eqref{pars}), meaning that the relation between diffusion and rotation is independent of the chemical state. The condition to attain a stationary chemotactic profile is (see SI)
\begin{equation}
\frac{v_d^{(C)}}{v_d^{(E)}} + \frac{e^{\Delta S_m} \mathcal{R}^{\rm eq} + \mathcal{M}(\vec{\mathscr{P}})}{b(\vec{\mathscr{P}}) e^{\Delta S_m} \mathcal{R}^{\rm eq} + c(\vec{\mathscr{P}})} < 1
\label{chemotaxis}
\end{equation}
Here, $\vec{\mathscr{P}}$ is a short notation for all free parameters involved. Notably, the condition \eqref{chemotaxis} holds independently of the presence of a catalytic step, even if the functions $\mathcal{M}$, $b$ and $c$ have to be modified according to the underlying chemical model. Their explicit expressions are provided in the SI. The inequality in Eq.~\eqref{chemotaxis} highlights the fact that, in order to observe steady-state chemotaxis, the enzyme has to diffuse faster than the complexes, being, for example, much smaller than the substrate. This is exactly the case of the experiment reported in \cite{guha}, where chemotaxis of molecular dyes is observed in polymer gradients.

\begin{figure}[t]
\includegraphics[width=\columnwidth]{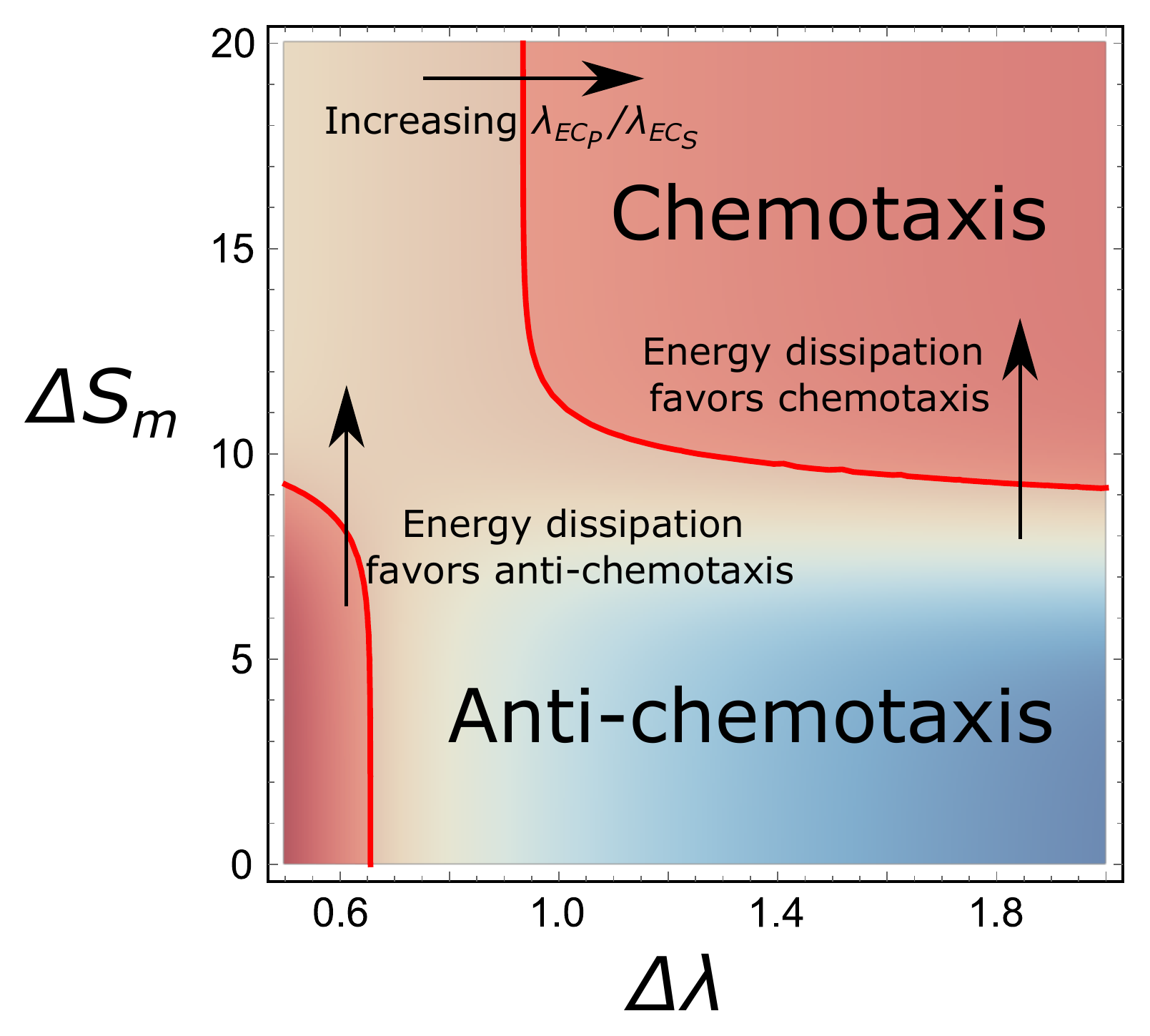}
\caption{Contour plot of the r.h.s. of Eq.~\eqref{chemotaxis} in the phase-space $(\Delta \lambda, \Delta S_m)$. Red shaded areas identify the region of parameters for which the system is chemotactic at stationarity. Intuitively, they are characterized by sizes of the kicks compensating the shortage of concentration (e.g. $[P] \ll [P]^{\rm eq}$ and $\lambda_{EC_P} \gg \lambda_{EC_S})$). Conversely, an anti-chemotactic behavior appears for $\Delta \lambda$ and $\Delta S_m$ lying in the remaining portion of space. The two vertical arrows sketch the direction of increasing dissipation for two different values of $\Delta \lambda$, while the horizontal arrow indicates the direction of increasing $\Delta \lambda$ for a fixed dissipation. Remarkably, energy dissipation can favor or disfavor the onset of molecular chemotaxis.}
\label{fig3}
\end{figure}

Here, we study an illustrative example in which $v_d^{(C)} = 2 v_d^{(E)}$, $\lambda_{EC_S} = \lambda_{C_SE} \neq \lambda_{C_PE} = \lambda_{EC_P}$, $\lambda_{C_SC_P} \ll \lambda_{EC_S}, \lambda_{EC_P}$, $\Delta S_m$ is free to vary, and all the other parameters are set to the fitted values for AChE (see Table \ref{table1}). In the SI, we show that the condition for the onset of chemotaxis is governed solely by the dissipated energy, $\Delta S_m$, and by
\begin{equation}
\Delta \lambda = \frac{1}{2} \left( 1 + \frac{\lambda_{EC_P}}{\lambda_{EC_S}} \right)
\label{dl}
\end{equation}
which quantifies the unbalance between substrate-induced kicks and product-induced kicks. If, for example, $\lambda_{EC_S} > \lambda_{EC_P}$ the main contribution from kicks during the catalytic cycle stems from the binding/unbinding of the substrate. It is worth noting that, if the catalytic step were absent, chemotaxis could still take place, provided a gradient of substrate concentration is present. For this particular case, we show in Fig.~\ref{fig3} the existence of regions of the phase-space $(\Delta \lambda, \Delta S_m)$ for which the system exhibits a chemotactic stationary profile. In simple terms, if $[S] \gg [S]^{\rm eq}$, i.e. in the strong dissipative regime, the enzyme tends to be steadily chemotactic when the product-induced kicks are much stronger than the substrate-induced ones, $\lambda_{EC_P} \gg \lambda_{EC_S}$, so to compensate the abundance of substrate with respect to the equilibrium value. Remarkably, depending on the value of $\Delta \lambda$, energy dissipation can favor chemotaxis, or be detrimental for it.


Our model identifies the core ingredients to effectively mimic the behavior of real enzyme molecular systems. The pursuit of building working nano or micro-motors exploiting enzymes' kinetics requires the ability to design systems for any desired purpose. In this perspective, it is crucial to have an intuitive grasp of the mechanisms acting at the molecular scale. Special importance has to be ascribed to the energy required to induce directed motion and enhanced diffusion, since these affect the potential efficiency of the machine. In this Letter, we proved that the injection of energy in the system can result in a switch of behavior, both from chemotaxis to anti-chemotaxis, and viceversa. Numerous experiments are needed to make up for the lack of systematic investigations in this direction.

A different, yet fascinating, perspective to be studied within the discussed framework is the emergence (or selection) of cellular chemotaxis, starting from an ensemble of self-replicating monomers. Anti-chemotactic molecules tend to move away from the substrate, being less abundant where, in contrast, chemotactic systems are in a great amount. The possibility of switching behavior changing few model parameters, e.g. only $\Delta \lambda$ in the example in Fig.~\ref{fig3}, at the same energy expenditure $\Delta S_m$, hints at the intriguing idea that repeated mutations may spontaneously select complex molecules exhibiting chemotaxis, as observed in bacteria \cite{chembac} and cells \cite{chemcell}.

\begin{widetext}

\clearpage

\section*{Supplementary Information}

\section{Microscopic model and continuum limit}

The microscopic model introduced in the main can be written for all possible states, $P_R^{(E)}$ being already reported (see Eq. (1) of the main text). We have:
\begin{eqnarray*}
P_R^{(E)}(x,t + \Delta t) &=& P_R^{(E)}(x,t) \Big( 1 - k^{(E)}_{\rm out} \Delta t \Big) + d \Delta t \left( (1-q) P_R^{(E)}(x - \Delta x) + q P_L^{(E)}(x - \Delta x) \right) + \\
&+& (1-q) \Delta t \Big( k_{-1} P_R^{(C_S)}\left( x - l_S \Delta x \right) + k_{2} P_R^{(C_P)}\left( x - l_P \Delta x \right) \Big) + \\
&+& q \Delta t \Big( k_{-1} P_L^{(C_S)}\left( x - l_S \Delta x \right) + k_{2} P_L^{(C_P)}\left( x - l_P \Delta x \right) \Big) \\
P_R^{(C_S)}(x,t + \Delta t) &=& P_R^{(C_S)}(x,t) \Big( 1 - k^{(C_S)}_{\rm out} \Delta t \Big) + d \Delta t \left( (1-q) P_R^{(C_S)}(x - \Delta x) + q P_L^{(C_S)}(x - \Delta x) \right) \\
&+& (1-q) \Delta t \Big( k_{D} [S(x - l_S \Delta x)] P_R^{(E)}\left( x - l_S \Delta x \right) + \kappa k_{C} P_R^{(C_P)}\left( x - l_C \Delta x \right) \Big) + \\
&+& q \Delta t \Big( k_{D} [S(x - l_S \Delta x)] P_L^{(C_S)}\left( x - l_S \Delta x \right) + \kappa k_{C} P_L^{(C_P)} \left( x - l_C \Delta x \right) \Big) \\
P_R^{(C_P)}(x,t + \Delta t) &=& P_R^{(C_P)}(x,t) \Big( 1 - k_{\rm out}^{(C_P)} \Delta t \Big) + d \Delta t \left( (1-q) P_R^{(C_P)}(x - \Delta x) + q P_L^{(C_P)}(x - \Delta x) \right) + \\
&+& (1-q) \Delta t \Big( k_{-2} [P(x - l_P \Delta x)] P_R^{(E)}\left( x - l_P \Delta x \right) + k_{C} P_R^{(C_S)}\left( x - l_C \Delta x \right) \Big) + \\
&+& q \Delta t \Big( k_{-2} [P(x - l_P \Delta x)] P_L^{(E)}\left( x - l_P \Delta x \right) + k_{C} P_L^{(C_S)}\left( x - l_C \Delta x \right) \Big) \nonumber \\
P_L^{(E)}(x,t + \Delta t) &=& P_L^{(E)}(x,t) \Big( 1 - k^{(E)}_{\rm out} \Delta t \Big) + d \Delta t \left( (1-q) P_L^{(E)}(x + \Delta x) + q P_R^{(E)}(x + \Delta x) \right) + \\
&+& (1-q) \Delta t \Big( k_{-1} P_L^{(C_S)}\left( x + l_S \Delta x \right) + k_{2} P_L^{(C_P)}\left( x + l_P \Delta x \right) \Big) + \\
&+& q \Delta t \Big( k_{-1} P_R^{(C_S)}\left( x + l_S \Delta x \right) + k_{2} P_R^{(C_P)}\left( x + l_P \Delta x \right) \Big) \\
P_L^{(C_S)}(x,t + \Delta t) &=& P_L^{(C_S)}(x,t) \Big( 1 - k^{(C_S)}_{\rm out} \Delta t \Big) + d \Delta t \left( (1-q) P_L^{(C_S)}(x + \Delta x) + q P_R^{(C_S)}(x + \Delta x) \right) \\
&+& (1-q) \Delta t \Big( k_{D} [S(x + l_S \Delta x)] P_L^{(E)}\left( x + l_S \Delta x \right) + \kappa k_{C} P_L^{(C_P)}\left( x + l_C \Delta x \right) \Big) + \\
&+& q \Delta t \Big( k_{D} [S(x + l_S \Delta x)] P_R^{(C_S)}\left( x + l_S \Delta x \right) + \kappa k_{C} P_R^{(C_P)} \left( x + l_C \Delta x \right) \Big) \\
P_L^{(C_P)}(x,t + \Delta t) &=& P_L^{(C_P)}(x,t) \Big( 1 - k_{\rm out}^{(C_P)} \Delta t \Big) + d \Delta t \left( (1-q) P_L^{(C_P)}(x + \Delta x) + q P_R^{(C_P)}(x + \Delta x) \right) + \\
&+& (1-q) \Delta t \Big( k_{-2} [P(x + l_P \Delta x)] P_L^{(E)}\left( x + l_P \Delta x \right) + k_{C} P_L^{(C_S)}\left( x + l_C \Delta x \right) \Big) + \\
&+& q \Delta t \Big( k_{-2} [P(x + l_P \Delta x)] P_R^{(E)}\left( x + l_P \Delta x \right) + k_{C} P_R^{(C_S)}\left( x + l_C \Delta x \right) \Big) \nonumber
\end{eqnarray*}
with:
\begin{eqnarray*}
k^{(E)}_{\rm out} = k_D [S(x)] + k_{-2} [P(x)] - 2 d \qquad
k^{(C_S)}_{\rm out} = k_{-1} + k_{C} - 2 d \qquad
k^{(C_P)}_{\rm out} = k_{2} + \kappa k_{C} - 2 d
\end{eqnarray*}
Dividing both sides by $\Delta t$, and performing the limit $\Delta t \to 0$, we recontruct the temporal derivative on the l.h.s.; hence, we expand the r.h.s. for $\Delta x \to 0$. As explained in the main text, we employ also the limit $q \to 0$, in order to introduce a persistence in the system, associated with the molecule orientation. Finally, we introduce the following finite parameters:
\begin{gather}
\label{constraints}
\lim_{\Delta x ,\Delta t\to 0} d \Delta x = v_d \quad
\lim_{q,\Delta t \to 0} q d = \tau_r^{-1} \\
\quad
\lim_{\Delta x,\Delta t \to 0} k_{XY} l_{XY} \Delta x = k_{XY} \lambda_{XY} \;\;\; \forall \; X,Y \quad
\lim_{q, \Delta t \to 0} q k_{XY} = k_{XY} \nu_{XY} \;\;\; \forall \; X,Y \nonumber
\end{gather}
meaning the all other combinations of small parameters must vanish in the limit $q \to 0$, $\Delta x \to 0$ and $\Delta t \to 0$. Here, $\Delta t$ is implicit in the definition of the rates. Notice that these parameters capture all information up to the leading non-zero order in the expansion. Substituting Eq. \eqref{constraints} after having expanded the microscopic model, we have:
\begin{eqnarray}
\label{2}
\partial_t P_R^{(E)} &=& - k_{-1} \lambda_{S} \partial_x P_R^{(C_S)} - k_2 \lambda_P \partial_x P_R^{(C_P)} - v_d \partial_x P_R^{(E)} + \nonumber \\
&-& k_{-1} \nu_{C_SE} \left( P_R^{(C_S)} - P_L^{(C_S)} \right) - k_2 \nu_{C_PE} \left( P_R^{(C_P)} - P_L^{(C_P)} \right) - \tau_r^{-1} \left( P_R^{(E)} - P_L^{(E)} \right) + \nonumber \\
&+& k_{-1} P_R^{(C_S)} + k_{2} P_R^{(C_P)} - \left( k_{D} [S(x)] + k_{-2} [P(x)] \right) P_R^{(E)}(x,t) \\
\partial_t P_R^{(C_S)} &=& - k_D \lambda_S \partial_x \left( [S(x)] P_R^{(E)} \right) - \kappa k_C \lambda_C \partial_x P_R^{(C_P)} - v_d \partial_x P_R^{(C_S)} + \nonumber \\
&-& k_D \nu_{EC_S} [S(x)] \left( P_R^{(E)} - P_L^{(E)} \right) - \kappa k_C \nu_{C_PC_S} \left( P_R^{(C_P)} - P_L^{(C_P)} \right) - \tau_r^{-1} \left( P_R^{(C_S)} - P_L^{(C_S)} \right) + \nonumber \\
&+& k_D [S(x)] P_R^{(E)} + \kappa k_C P_R^{(C_P)} - \left( k_{-1} + k_C \right) P_R^{(C_S)}(x,t) \\
\partial_t P_R^{(C_P)} &=& - k_{-2} \lambda_S \partial_x \left( [P(x)] P_R^{(E)} \right) - k_C \lambda_C \partial_x P_R^{(C_S)} - v_d \partial_x P_R^{(C_P)} + \nonumber \\
&-& k_{-2} \nu_{EC_P} [P(x)] \left( P_R^{(E)} - P_L^{(E)} \right) - k_C \nu_{C_SC_P} \left( P_R^{(C_S)} - P_L^{(C_S)} \right) - \tau_r^{-1} \left( P_R^{(C_P)} - P_L^{(C_P)} \right) + \nonumber \\
&+& k_{-2} [P(x)] P_R^{(E)} + k_C P_R^{(C_S)} - \left( k_{2} + \kappa k_C \right) P_R^{(C_P)}(x,t) \\
\partial_t P_L^{(E)} &=& k_{-1} \lambda_{S} \partial_x P_L^{(C_S)} + k_2 \lambda_P \partial_x P_L^{(C_P)} + v_d \partial_x P_L^{(E)} + \nonumber \\
&-& k_{-1} \nu_{C_SE} \left( P_L^{(C_S)} - P_R^{(C_S)} \right) - k_2 \nu_{C_PE} \left( P_L^{(C_P)} - P_R^{(C_P)} \right) - \tau_r^{-1} \left( P_L^{(E)} - P_R^{(E)} \right) + \nonumber \\
&+& k_{-1} P_L^{(C_S)} + k_{2} P_L^{(C_P)} - \left( k_{D} [S(x)] + k_{-2} [P(x)] \right) P_L^{(E)}(x,t) \\
\partial_t P_L^{(C_S)} &=& k_D \lambda_S \partial_x \left( [S(x)] P_L^{(E)} \right) + \kappa k_C \lambda_C \partial_x P_L^{(C_P)} + v_d \partial_x P_L^{(C_S)} + \nonumber \\
&-& k_D \nu_{EC_S} [S(x)] \left( P_L^{(E)} - P_R^{(E)} \right) - \kappa k_C \nu_{C_PC_S} \left( P_L^{(C_P)} - P_R^{(C_P)} \right) - \tau_r^{-1} \left( P_L^{(C_S)} - P_R^{(C_S)} \right) + \nonumber \\
&+& k_D [S(x)] P_L^{(E)} + \kappa k_C P_L^{(C_P)} - \left( k_{-1} + k_C \right) P_L^{(C_S)}(x,t) \\
\partial_t P_L^{(C_P)} &=& k_{-2} \lambda_S \partial_x \left( [P(x)] P_L^{(E)} \right) + k_C \lambda_C \partial_x P_L^{(C_S)} + v_d \partial_x P_L^{(C_P)} + \nonumber \\
&-& k_{-2} \nu_{EC_P} [P(x)] \left( P_L^{(E)} - P_R^{(E)} \right) - k_C \nu_{C_SC_P} \left( P_L^{(C_S)} - P_R^{(C_S)} \right) - \tau_r^{-1} \left( P_L^{(C_P)} - P_R^{(C_P)} \right) + \nonumber \\
&+& k_{-2} [P(x)] P_L^{(E)} + k_C P_L^{(C_S)} - \left( k_{2} + \kappa k_C \right) P_L^{(C_P)}(x,t)
\label{7}
\end{eqnarray}
where, due to the scallop theorem \cite{purcell}, $\lambda_{EC_S} = \lambda_{C_SE} \equiv \lambda_S$, $\lambda_{EC_P} = \lambda_{C_PE} \equiv \lambda_P$, and $\lambda_{C_SC_P} = \lambda_{C_PC_S} \equiv \lambda_C$.

The chemical part of the model consists in two contributions: one local and the other non-local, due to reaction kicks. We have already seen that the latter is finite in the continuum limit, i.e. $k_{XY} l_{XY}\Delta x \to k_{XY} \lambda_{XY}$. As a consequence, the former, which is just $k_{XY}$ must scale as $(\Delta x)^{-1}$. In the limit of small $\Delta x$, we cure this mathematical (not physical) divergence by employing a fast time-scale separation. It corresponds to the following substitution for the probability distribution:
\begin{equation}
P_{R,L}^{(X)}(x,t) = \pi^{(X)}(x)~p_{R,L}(x,t)
\label{fast}
\end{equation}
where $\pi^{(X)}$ is the stationary solution of the chemical system (without kicks) for the state $X$. Equivalently, we are saying that the local chemistry converges much rapidly than all other processes to a steady state. Substituting Eq. \eqref{fast} into Eqs. \eqref{2}-\eqref{7}, we obtain:
\begin{eqnarray}
\label{pr}
\partial_t p_R &=& - \partial_x \big( \langle L(x) \rangle p_R \big) - \langle N(x) \rangle \left( p_R - p_L \right) \\
\partial_t p_L &=& \partial_x \big( \langle L(x) \rangle p_L \big) + \langle N(x) \rangle \left( p_R - p_L \right)
\label{pl}
\end{eqnarray}
where
\begin{gather*}
\langle L \rangle = \sum_{XY} k_{XY} \lambda_Y \pi^{(X)} + v_d \qquad \qquad
\langle N \rangle = \sum_{XY} k_{XY} \nu_Y \pi^{(X)} + \tau_r^{-1}
\end{gather*}
We derive $p_L$ from Eq. \eqref{pr}, and substitute it into Eq. \eqref{pl}. Doing the opposite procedure for $p_R$, we get:
\begin{eqnarray}
\left( \partial_t - \partial_x \langle L \rangle \right) \frac{1}{\langle N \rangle} \left( \partial_t + \partial_x \langle L \rangle \right) p_R + 2 \partial_t p_R &=& 0 \nonumber \\
\left( \partial_t + \partial_x \langle L \rangle \right) \frac{1}{\langle N \rangle} \left( \partial_t - \partial_x \langle L \rangle \right) p_L + 2 \partial_t p_L &=& 0
\label{eq4main}
\label{teq}
\end{eqnarray}
The set of equation \eqref{eq4main} corresponds to Eq. (4) of the main text.

Since we are interested in the total probability of finding the enzyme in any state with any orientation, $P = p_R + p_L$, summing the two equations above, we obtain:
\begin{gather}
\partial_{tt} P(x,t) + 2 \langle N(x) \rangle ~\partial_t P(x,t) - \langle N(x) \rangle ~\partial_x \bigg( \frac{\langle L(x) \rangle}{\langle N(x) \rangle} \partial_x \left( \langle L(x) \rangle P(x,t) \right) \bigg) + \nonumber \\
+ \bigg( \langle N(x) \rangle ~\partial_x \left( \frac{\langle L(x) \rangle}{\langle N(x) \rangle} \partial_t P(x,t) \right) - \partial_x \left( \langle L(x) \rangle ~\partial_t P(x,t) \right) \bigg) = 0
\label{eqgen}
\end{gather}
At stationarity, this equation reduces to Eq. (5) of the main text:
\begin{equation}
\langle N(x) \rangle ~\partial_x \bigg( \frac{\langle L(x) \rangle}{\langle N(x) \rangle} \partial_x \left( \langle L(x) \rangle P(x,t) \right) \bigg) = 0
\end{equation}
whose solution is:
\begin{equation}
P^{\rm ss} = \mathscr{N} \frac{1}{\langle L(x) \rangle}
\end{equation}
where $\mathscr{N}$ is a normalization factor.

\section{Free parameters for the fitting procedure}

Let us start deriving the functional form of $\langle L \rangle$:
\begin{eqnarray}
\langle L \rangle &=& v_d \left( 1 + \frac{ \alpha_S \left( 1 + \mu_S \right) [S(x)]}{\mathcal{K}_S \left( \mu_S + \gamma_S \right) + [S(x)] \left( \frac{k_2}{\bar{k}_S} \mu_S + 1 + \kappa \right) + \frac{k_{-2}}{k_D} [P(x)] \left( \frac{k_{-1}}{\bar{k}_S} \mu_S + 1 + \kappa \right)} \right. + \nonumber \\
&+& \left. \frac{ \alpha_P \left( 1 + \mu_P \right) [P(x)]}{\mathcal{K}_P \left( \mu_P + \gamma_P \right) + [S(x)] \left( \frac{k_2}{\bar{k}_P} \mu_P + 1 + \kappa \right) + \frac{k_{-2}}{k_D} [P(x)] \left( \frac{k_{-1}}{\bar{k}_P} \mu_P + 1 + \kappa \right)} \right)
\label{eqL}
\end{eqnarray}
where:
\begin{gather}
\alpha_S = \frac{\lambda_S}{\lambda_0} \;\;\;\;\;\;\; \lambda_0 = \frac{v_d}{2 k_D \mathcal{K}_S} \;\;\;\;\;\;\; \mathcal{K}_S = \frac{\zeta_S}{k_D} \;\;\;\;\;\;\; \gamma_S = \frac{k_{-1} \kappa + k_2}{\zeta_S} \nonumber \\
\zeta_S = k_{-1} \kappa + \lambda_{PC} k_2 + \lambda_{CS} \kappa k_C \;\;\;\;\;\;\; \mu_S = \frac{\bar{k}_S}{k_C} \;\;\;\;\;\;\; \bar{k}_S = \frac{k_{-1} k_2}{\zeta_S} \nonumber \\
\lambda_{PC} = \frac{1}{2} \left( 1 + \frac{\lambda_P + \lambda_C}{\lambda_S}\right) \;\;\;\;\;\;\; \lambda_{CS} = \frac{\lambda_{C}}{\lambda_S}
\label{p1}
\end{gather}
and, analogously:
\begin{gather}
\alpha_P = \frac{\lambda_P}{\lambda_0^{(P)}} \;\;\;\;\;\;\; \lambda_0^{(P)} = \frac{v_d}{2 k_{-2} \mathcal{K}_P} \;\;\;\;\;\;\; \mathcal{K}_P = \frac{\zeta_P}{k_D} \;\;\;\;\;\;\; \gamma_P = \frac{k_{-1} \kappa + k_2}{\zeta_P} \nonumber \\
\zeta_P = k_{-1} \kappa \lambda_{SC} + k_2 + \lambda_{CP} \kappa k_C \;\;\;\;\;\;\; \mu_P = \frac{\bar{k}_P}{k_C} \;\;\;\;\;\;\; \bar{k}_P = \frac{k_{-1} k_2}{\zeta_P} \nonumber \\
\lambda_{SC} = \frac{1}{2} \left( 1 + \frac{\lambda_S + \lambda_C}{\lambda_P}\right) \;\;\;\;\;\;\; \lambda_{CP} = \frac{\lambda_{C}}{\lambda_P}
\end{gather}

However, it can be proven that the denominators of the terms in Eq. \eqref{eqL} are the same. Hence, we can write the following formula:
\begin{equation}
\langle L \rangle = v_d \left( 1 + \frac{[S(x)] + \mathcal{M} [P(x)]}{a + b [S(x)] + c [P(x)]} \right)
\label{Lfit}
\end{equation}
with:
\begin{equation}
a = \frac{\mathcal{K}_S \left( \mu_S + \gamma_S \right)}{\alpha_S \left( 1 + \mu_S \right)} \;\;\;\;\;\;\; b = \frac{\frac{k_2}{\bar{k}_S} \mu_S + 1 + \kappa}{\alpha_S \left( 1 + \mu_S \right)} \;\;\;\;\;\;\; c = \frac{k_{-2}}{k_D} \frac{\frac{k_{-1}}{\bar{k}_S} \mu_S + 1 + \kappa}{\alpha_S \left( 1 + \mu_S \right)}
\label{p2}
\end{equation}
and the proportionality factor is:
\begin{equation}
\mathcal{M} = \frac{\alpha_P}{\alpha_S} \frac{\mathcal{K}_S}{\mathcal{K}_P} \frac{\left( 1 + \mu_P \right)}{\left( 1 + \mu_S \right)} \frac{\mu_S}{\mu_P}
\label{p3}
\end{equation}
Introducing the information about energy dissipation:
\begin{equation}
\frac{[S(x)]}{[P(x)]} = e^{\Delta S_m} \frac{[S]^{\rm eq}}{[P]^{\rm eq}} = e^{\Delta S_m} \mathcal{R}^{\rm eq}
\label{energy}
\end{equation}
Eq. \eqref{Lfit} can be rewritten as follows:
\begin{equation}
\langle L \rangle = v_d \left( 1 + \frac{e^{\Delta S_m} \mathcal{R}^{\rm eq} + \mathcal{M}}{\frac{a}{[S]} e^{\Delta S_m} \mathcal{R}^{\rm eq} + b e^{\Delta S_m} \mathcal{R}^{\rm eq} + c} \right)
\end{equation}
Since $k_2 \gg k_{-1}, k_C$, $\kappa$ is expected to be less than unity, $\lambda_{PC}, \lambda_{CS}, \lambda_{SC}$ and $\lambda_{CP}$ have to be $\mathcal{O}(1)$ because they account for the unbalance of the kicks associated with different chemical reactions, and guessing, for simplicity, that $\lambda_{S} = \lambda_C = \lambda_P \equiv \lambda$, we can rewrite $\langle L(x) \rangle$ in the following approximate form:
\begin{equation}
\langle L \rangle = v_d \left( 1 + \frac{e^{\Delta S_m} \frac{k_{-2} \kappa k_{-1}}{k_D k_2} + \frac{2 + 2 \kappa + 3 k_{-1}/k_C}{1 + k_{-1}/k_C}\frac{2 k_{-2}}{3 k_D}}{\frac{2 k_2 k_D + 3 k_2 k_{-1}}{2 k_D^2 \alpha_S + 2 k_D k_{-1} \alpha_S} \frac{1}{[S]} e^{\Delta S_m} \frac{k_{-2} \kappa k_{-1}}{k_D k_2} + \frac{3 k_2 + 2 k_D + 2 \kappa k_D}{2 k_D \alpha_S + 2 k_{-1} \alpha_S} \left( e^{\Delta S_m} \frac{k_{-2} \kappa k_{-1}}{k_D k_2} + \frac{k_{-2}}{k_D} \frac{3 k_{-1} + 2 k_D (1 + \kappa)}{3 k_2 + 2 k_D (1 + \kappa)} \right)} \right)
\label{eqapp}
\end{equation}
For brevity of notation, we write: $\langle L \rangle = v_d \left( 1 + \mathcal{L}(\vec{\mathscr{P}}) \right)$. The advantage of using Eq. \eqref{eqapp} to perform the fitting procedure is that it depends on a much smaller number of free parameters with respect to the full expression. Clearly, all fitted values have to be compatible with the introduced approximations, as shown in the main text.

An analogous procedure can be performed on $\langle N(x) \rangle$. We obtain that the expression of the effective diffusion coefficient (see Eq. (6) of the main text) is:
\begin{equation}
D^{\rm eff} = D_0 \left( 1 + \mathcal{L}(\vec{\mathscr{P}}) \right)^2 \left( 1 + r ~\mathcal{L}(\vec{\mathscr{P}}) \right)^{-1}
\label{dfit}
\end{equation}
where we have defined the following quantities:
\begin{equation}
r = \frac{\nu}{\lambda} v_d \tau_r \qquad \qquad D_0 = \frac{v_d^2 \tau_r}{2}
\end{equation}
Once we fixed $k_2$, $k_D$ and $k_{-1}$ to reasonable values (see Eq. (8) of the main text), Eqs. \eqref{eqapp} and \eqref{dfit} depend only on the free parameters discussed in the main text.

It is worth noting that the expressions in Eqs.~\eqref{eqapp} and \eqref{dfit} do not explicitly depend on $k_{-2}$, which remains a floating parameter within the fitting procedure.

\subsection{Important remarks on the fitting procedure}

In order to simultaneously fit the profiles of both enzyme concentration and effective diffusion coefficient, we minimize the following function with respect to the free parameters:
\begin{equation}
\mathcal{F} = \sum_{i \in {\rm data}} \left( \frac{\left( P^{\rm ss}(x_i) - P^{\rm exp}(x_i) \right)^2}{|\epsilon_i|} + \zeta \frac{\left( D^{\rm eff}(x_i) - D^{\rm exp}(x_i) \right)^2}{|\eta_i|} \right)
\end{equation}
where $P^{ss}(x_i)$ and $P^{\rm exp}(x_i)$ are theoretical and experimental value of the stationary enzyme profile at the point $x_i$, $D^{\rm exp}(x_i)$ is the measured effective diffusion coefficient at $x_i$, and $\epsilon_i$ and $\eta_i$ are errors on profiles of concentration and diffusion coefficient, respectively. $\zeta$ is a weight parameter tuned to optimize the fit. The sum is performed over all available data points. We also impose hard constraints on the model-parameters, since they must lie within physically reasonable ranges. Moreover, the set of parameters we obtain is not unique, in the sense that several similar choices may lead to similar results. Indeed, we stress the fact that the values of fitted model-parameters have to be intended as an order of magnitude, rather than precise predictions.

Another important point to discuss concerns the value of the entropy change, determining how far the system is from equilibrium. In our fitting procedure we always look for the smallest possible $\Delta S_m$ such that the system exhibits the measured behavior. This is in agreement with the idea that it is unlikely for a molecular system to have additional energy consumption without any advantage in terms of diffusion or taxis.

\section{Fitting procedure assuming only heat-induced kicks}

In this section, we perform the fit for the profiles of both concentration and effective diffusion coefficient for aceticholinesterase (AChE) and urease employing the assumption that only heat-induced kicks are present, i.e. $\lambda_S = \lambda_P = 0$. Heat-induced kicks are generated along with the catalysis of the substrate, associated with the rates $k_C$ and $\kappa k_C$, and the size $\lambda_C$. Following the derivation detailed in the section above, the resulting equations are:
\begin{equation*}
\langle L \rangle^{\rm heat} = v_d \left( 1 + \frac{\alpha_C  k_D \left( e^{\Delta S_m} \mathcal{R}^{\rm eq} \left( 2 \kappa k_C + k_2 \right) + \frac{k_{-2}}{k_D} \left( 2 \kappa k_C + \kappa k_{-1} \right) \right)}{\frac{e^{\Delta S_m} \mathcal{R}^{\rm eq}}{[S(x)]} \left( k_{-1} \kappa k_C + k_C k_2 + k_{-1} k_2 \right) + e^{\Delta S_m} \mathcal{R}^{\rm eq} \left( k_C (1 + \kappa) + k_2 \right) k_D + k_{-2} \left( k_C (1 + \kappa) + k_{-1} \right) k_C} \right)
\end{equation*}
where we introduced
\begin{equation}
\alpha_C = 2 \frac{\lambda_C}{v_d} k_C
\end{equation}
as the version of the Damk$\ddot{\rm o}$ler number in the presence of heat-induced kicks only, that we expect to be approximately unity. On the contrary, $\langle N \rangle$ is the same as before, since the change of orientation still affects all chemical states. Hence, for the effective diffusion coefficient, we have:
\begin{equation}
D^{\rm eff} = D_0 \Big( \langle L \rangle^{\rm heat} \Big)^2 \left( 1 + \mathcal{L}_\nu(\vec{\mathscr{P}})\right)^{-1}
\label{dheat}
\end{equation}
where $\mathcal{L}_\nu$ is equal to $\mathcal{L}$ defined above, with the replacement $\alpha_S \to \nu$, assuming that the rotation rate is equal for all states. The fitting parameter in this case is: $\mathcal{N} = \nu/\nu_0$, where $\nu_0 = (2 \tau_r k_2)^{-1}$. Again, $\mathcal{N}$ is a version of the Damk$\ddot{\rm o}$hler number for the rotation mechanism.

\begin{figure}[h]
\includegraphics[width=0.55 \columnwidth]{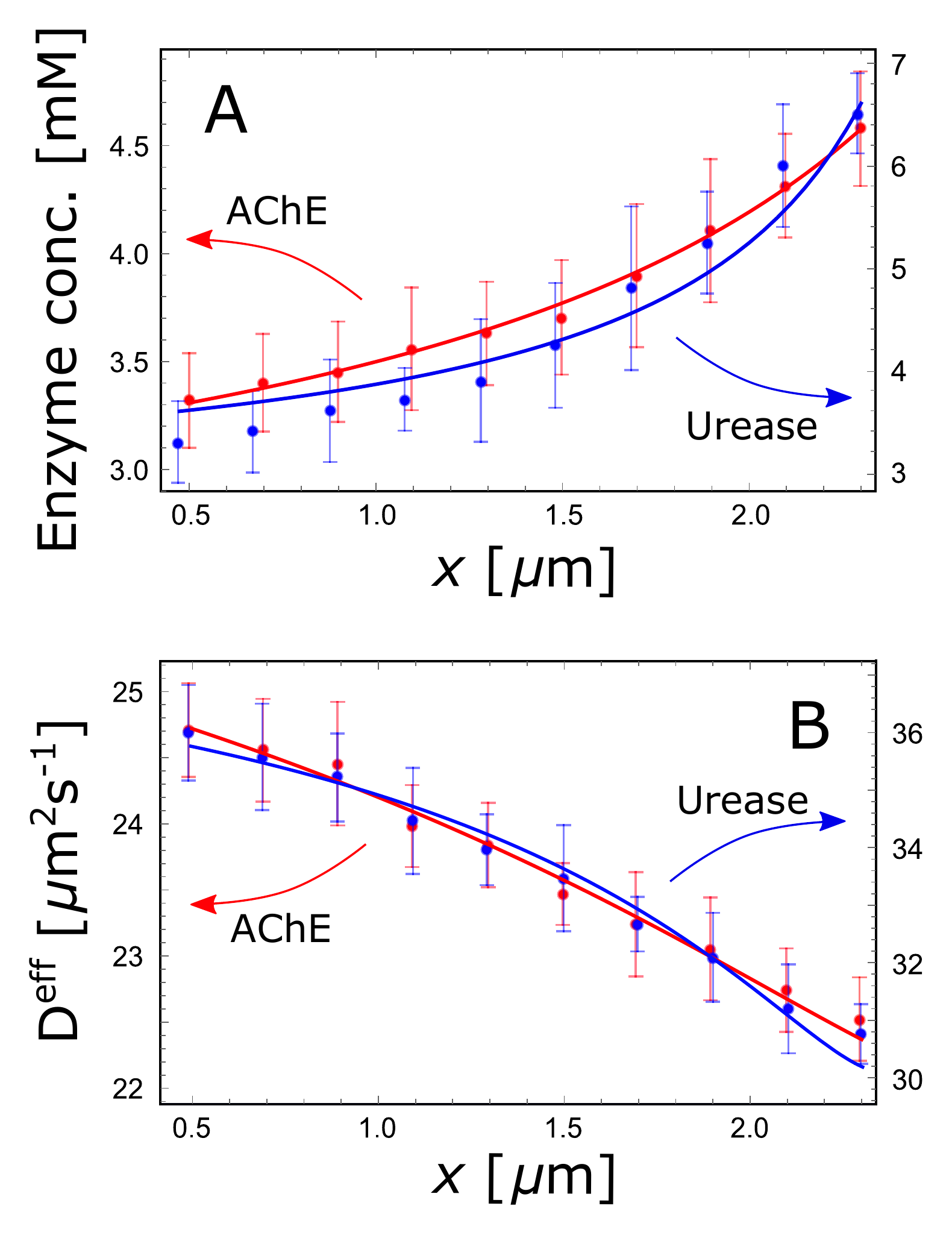}
\caption{\textit{Panel A} - Comparison between data extracted from \cite{granick} (dots) and theoretical predictions (line) for the stationary concentration profile of acetylcholinesterase (AChE) (in red) and urease (in blue) in the presence of heat-induced kicks only. Vertical bars indicate the experimental error. \textit{Panel B} - Comparison between data (dots) and theory, Eq.~\eqref{dheat} (line) for the profile of effective diffusion coefficient of AChE (in red) and urease (in blue).}
\label{SI1}
\end{figure}

\begin{table}
\begin{tabular}{ccc}
\hline
\;\;\; \textbf{Enzyme} \;\;\;  & \;\;\; \textbf{Parameter} \;\;\; & \;\;\; \textbf{Approximate Value} \\
\hline
\\
AChE & $D_0$ & 22.2 $\mu m^2 s^{-1}$ \\
         & $\Delta S_m$ & $16.4$ \\
         & $\kappa$ & $0.05$ \\
         & $k_C$ & $3.7 \times 10^3 ~s^{-1}$ \\
         & $\alpha_C$ & $1.6$ \\
         & $\mathcal{N}$ & $8.4$ \\
\\       
Urease & $D_0$ & $30.6 ~\mu m^2 s^{-1}$ \\
           & $\Delta S_m$ & $18.3$ \\
           & $\kappa$ & $0.05$ \\
           & $k_C$ & $2.4 \times 10^4 ~s^{-1}$ \\
           & $\alpha_C$ & $6.0$ \\
           & $\mathcal{N}$ & $10.0$
\end{tabular}
\caption{List of parameters with their approximate value. Chemical rates and model-dependent parameters are compatible with physical expectations. The bare diffusion coefficient $D_0$ lies in the range of measured values, according to \cite{granick}.}
\label{SIT}
\end{table}

This setting is not a further approximation with respect to the case presented in the main text, but it exploits different constraints on the size of the kicks. The fits are striking also in this case, as shown in Fig. \ref{SI1}, and all fitted values lie within physical ranges (see Table \ref{SIT}). An observation, compatible with experimental results, is that $\mathcal{N}$, which is related to the inverse of the turnover rate, is higher for Urease than for AChE, implying that AChE has a faster turnover than Urease. We conclude that our model is robust with respect to different \textit{reasonable} choices for the kicks. As explained in the main text, further experiments are needed to elucidate what is the correct microscopic assumption, in particular concerning what happens for non-catalytic molecules.

\section{The role of catalysis}

Here, we investigate the role of catalysis in explaining the data reported in \cite{granick}. We show that, if we neglect hydrolysis and synthesis of the substrate, i.e. $k_C = 0$, the model substantially fail in reconstructing the experiments, meaning that the catalytic step is indeed a crucial ingredient. However, it is worth noting that an anti-chemotactic behavior, along with an enhanced diffusion profile, could still take place in the presence of a gradient of substrate concentration. Further experiments could shed some light on these theoretical predictions about molecules that do not catalyze the substrate-to-product conversion. 

In this setting, we have the following expressions:
\begin{gather}
\langle L \rangle^{\rm /cat} = 1 + \frac{\alpha_S \left( \frac{k_{-2}}{k_D} \frac{\lambda_P}{\lambda_S} + e^{\Delta S_m} \mathcal{R}^{\rm eq} \right)}{\frac{k_{-1}}{k_D} \frac{e^{\Delta S_m} \mathcal{R}^{\rm eq}}{[S(x)]} + e^{\Delta S_m} \mathcal{R}^{\rm eq} + \frac{k_{-1} k_{-2}}{k_2 k_D}} = 1 + \mathcal{L}^{\rm /cat} \nonumber \\
D^{\rm eff} = D_0 \left( 1 + \mathcal{L}^{\rm /cat} \right)^2 \left( 1 + r ~\mathcal{L}^{\rm /cat} \right)^{-1}
\end{gather}

In Fig. \ref{SI2}, we show a comparison between the results of the fitting procedure with and without catalysis. As said above, neglecting the catalytic step leads to a stark worsening of the fits, albeit the onset of an anti-chemotactic profile seems a possible \textit{predictable} effect even for non-catalytic molecules. We do not report the fitted values in this case, since they do not contain any information about physical parameters.

\begin{figure}[h]
\includegraphics[width=\columnwidth]{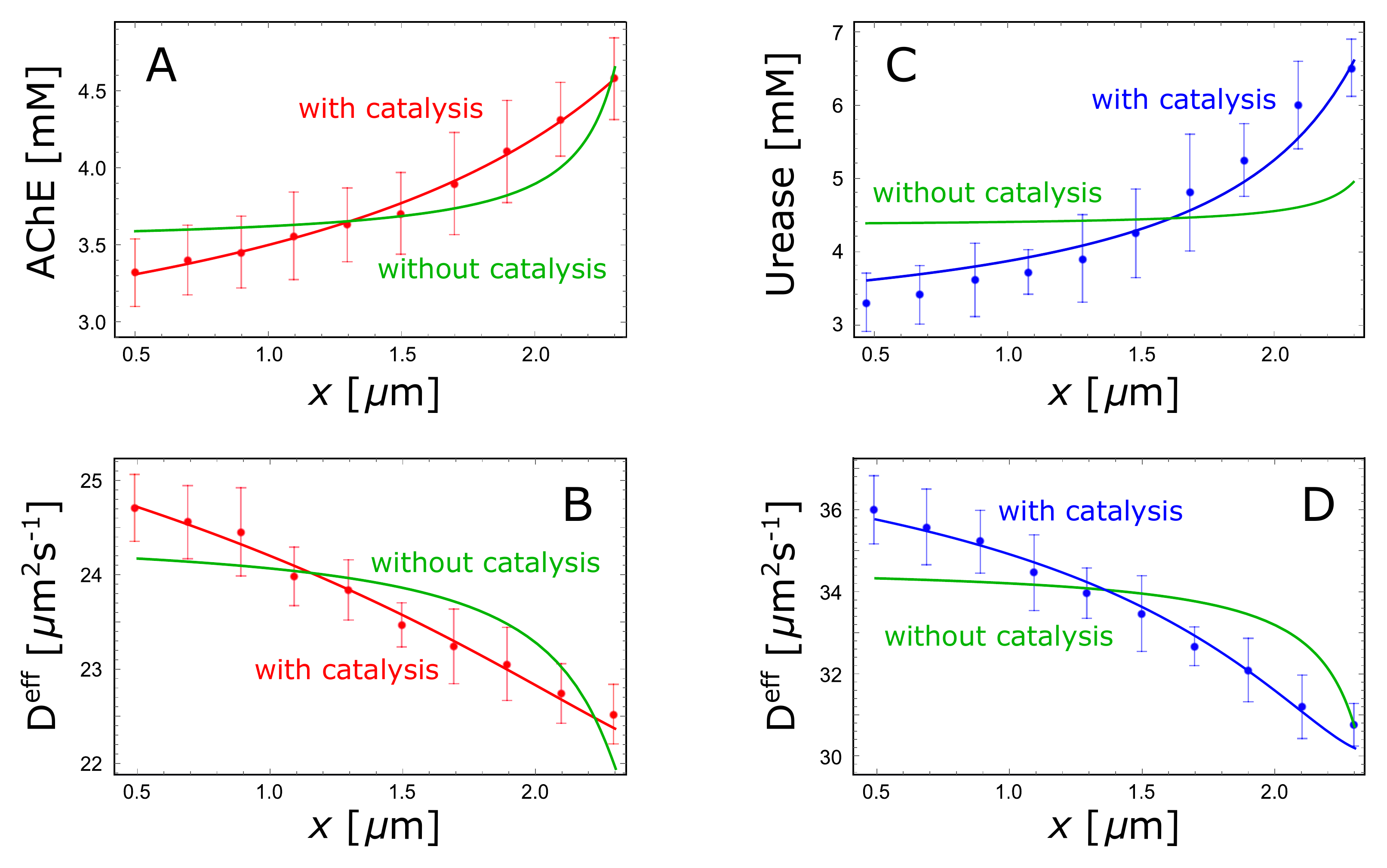}
\caption{\textit{Panel A} - Comparison between data extracted from \cite{granick} (dots), theoretical predictions in the presence (red line) and in the absence (green line) of catalysis for the stationary concentration profile of acetylcholinesterase (AChE). Vertical bars indicate the experimental error. \textit{Panel B} - The comparison is presented for the profile of the effective diffusion coefficient of AChE with the same color-code as for \textit{Panel A}. \textit{Panel C} - Dots represent data from \cite{granick}, with their error (vertical bars), for Urease. The blue line is the fit in the presence of catalysis, whereas the green line represents the fit without catalysis. \textit{Panel D} - The comparison is shown for the effective diffusion coefficient of Urease with the same color-code as for \textit{Panel C}.}
\label{SI2}
\end{figure}

\section{Short-time ballistic regime}

We have discussed, in the main text, the presence of a ballistic-to-diffusive transition in the simple case of a uniform substrate concentration. However, introducing the gradient $[S(x)]$, we can study both the short- and the long-time limit. In the long-time limit, $t \gg \left( 2 \langle N \rangle \right)^{-1}$, we have:
\begin{equation}
\partial_{t} P(x,t) - \partial_x \left( \frac{\langle L(x) \rangle}{2 \langle N(x) \rangle} \partial_x \left( \langle L(x) \rangle P(x,t) \right) \right) = 0
\end{equation}
An effective diffusion coefficient can be identified, as explained in the main text, in the limit of fast local chemical reactions, i.e. local chemical stationarity.

In order to investigate the short-time limit, we employ a numerical simulation of the complete dynamical equation, Eq. \eqref{eqgen}. In Fig. \ref{SI3} we show that a ballistic regime at initial stages of the dynamics is observed through its fingerprint, i.e. a quadratic dependence on time of the second spatial moment. In formulas, $\langle x^2 \rangle \propto t^2$ denotes a ballistic motion. A quantitative analysis, however, it is not possible in this context, since $v_d$ and $\tau_r$ are not known. We do not report numbers in Fig. \ref{SI3} since they are only a consequence of arbitrary choices we made on these latter parameters, and do not provide an informative result.

\begin{figure}[h]
\includegraphics[width=0.5 \columnwidth]{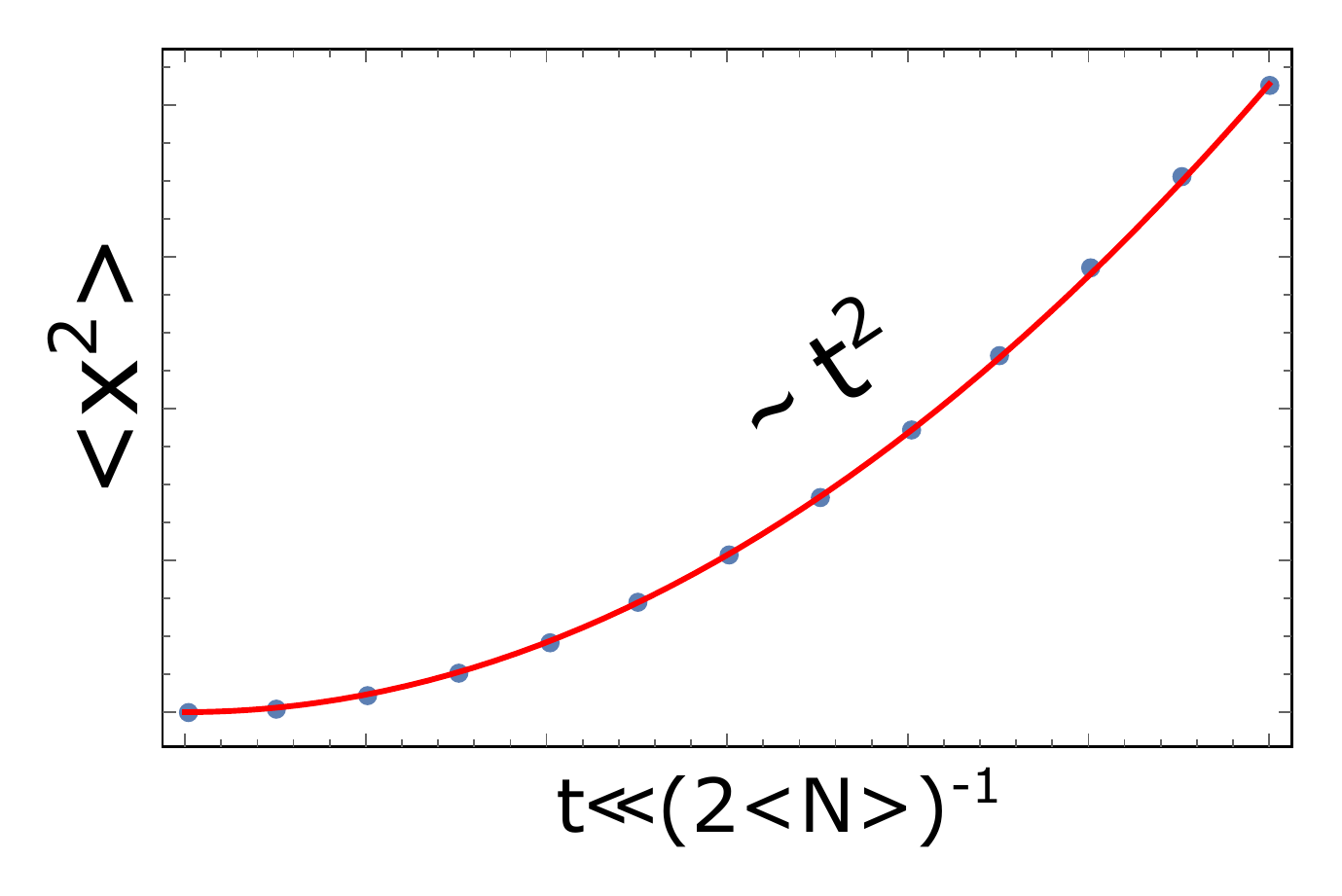}
\caption{Short-time behavior of the second moment $\langle x^2 \rangle$ as a function of time for a choice of parameters compatible with the fitted values for AChE. $\tau_r$ has been arbitrarily set to $10^{-3}$ for sake of simplicity, and $v_d$ so that $v_d^2 ~\tau_r = 2 D_0$. Points are data extracted from simulations, and the line is just a guide for the eye.}
\label{SI3}
\end{figure}

\section{Transient chemotactic behavior}

In the main text we extensively discussed the setting of the experiment reported in \cite{sen}. The authors observe a transient chemotactic behavior of the enzyme. Since the diffusion of the substrate is much faster than the one of the enzyme, it is reasonable to think that the enzyme will be affected, effectively, by a uniform substrate concentration. Moreover, in broad terms the time necessary for the substrate to reach a uniform distribution is larger than the one for the mixed system (enzyme + substrate) to start exhibiting a diffusive behavior. In fact,
\begin{equation}
\frac{L^2}{D^{\rm sub}} \approx \frac{200^2 \mu m^2}{\alpha 30.6 ~\mu m^2 s^{-1}} > \frac{1}{2 \langle N \rangle} \approx \frac{\tau_r}{4} \Rightarrow \alpha \tau_r < 5.2 \times 10^3 ~s
\label{comparison}
\end{equation}
where $\alpha$ quantifies the discrepancy between substrate and enzyme diffusion coefficient ($\alpha \ll 1$). All parameters have been fixed compatibly to the fitted values for Urease, albeit the value of the substrate for which they have been found is $10^3$ times lower than the one used in \cite{sen}. Since $\tau_r$ is expected to be comparable to the inverse of the turnover rate, $[10^{-3},10^{-4}]$, Eq. \eqref{comparison} is likely to be satisfied in the considered setting.

Hence, we study the dynamics of the system (substrate + enzyme) in the diffusive regime, with a uniform substrate concentration, starting with the enzyme entirely located in half of the capillary. The profile of enzyme concentration (in any chemical state) after an \textit{arbitrary} short time is shown in Fig. \ref{SIsen}. It is qualitatively similar to the one presented in \cite{sen}, meaning that this transient chemotactic behavior is the logical consequence of the initial conditions for a system \textit{en route} to a stationary flat distribution with an effective enhanced diffusion coefficient.

\begin{figure}[t]
\includegraphics[width=\columnwidth]{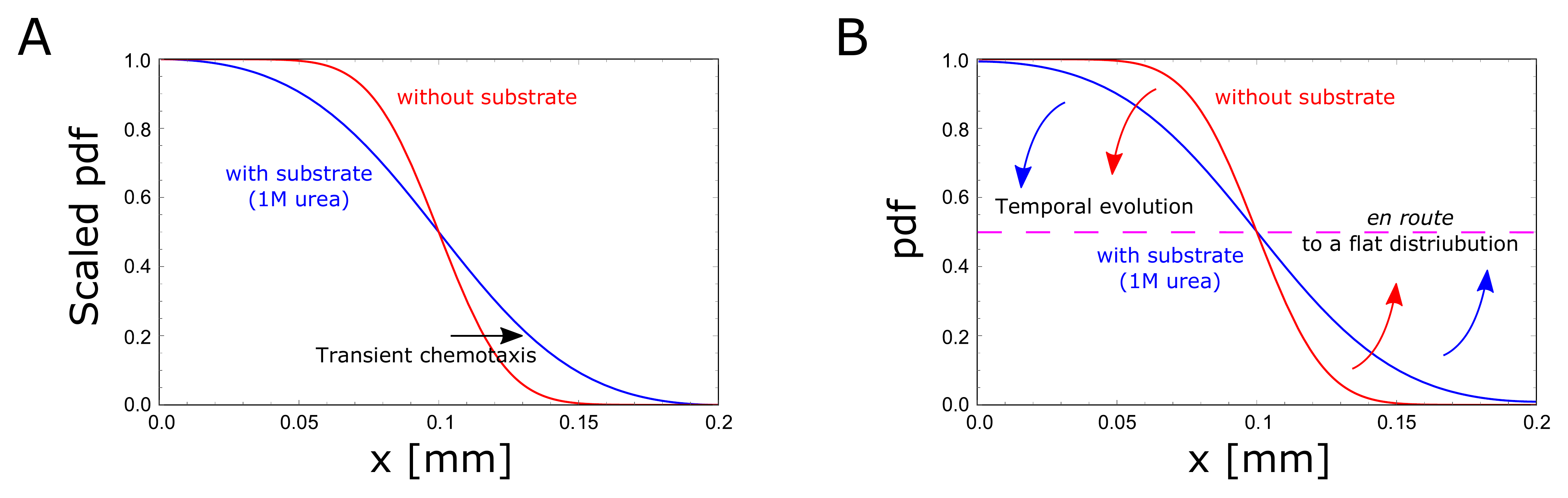}
\caption{\textit{Panel A} - Scaled probability distribution of the enzyme, such that it lies between $0$ and $1$, at short times. All parameters are in agreement with the condition \eqref{comparison}. \textit{Panel B} - Probability density functions (not scaled but properly normalized) for the same cases of \textit{Panel A} are shown (red and blue curves). The magenta dashed line represents the flat stationary distribution to which both systems (with and without substrate) eventually converge.}
\label{SIsen}
\end{figure}

\section{The onset of a stationary chemotactic profile in a simple example}

Consider the case in which the complexes have the same diffusion rate: $v_d^{(C_S)} = v_d^{(C_P)} = v_d^{(C)}$. In general, $\langle L \rangle$ is defined as the following average over the stationary distribution of chemical states (see main text):
\begin{equation}
\langle L \rangle = \sum_{XY} k_{XY} \lambda_{XY} \pi^{(X)} + \sum_X v_d^{(X)} \pi^{(X)} = \left( v_d^{(E)} \pi^{(E)} + v_d^{(C)} \left( \pi^{(S)} + \pi^{(P)} \right)  + \frac{[S(x)] + \mathcal{M} [P(x)]}{a + b [S(x)] + c [P(x)]} \right)
\label{eq32}
\end{equation}
It reduces to Eq.~\eqref{Lfit} when $v_d^{(E)} = v_d^{(C)} = v_d$, since $\pi^{(E)} + \pi^{(S)} + \pi^{(P)} = 1$ by construction. In this case, we notice that $D_0$ has to be defined by the following relation:
\begin{equation}
(v_d^{(E)})^2 \tau_r = 2 D_0
\end{equation}
since $v_d^{(E)}$ is the only diffusive rates that does not depend on the substrate, as $D_0$. Starting from Eq.~\eqref{eq32}, dividing and multiplying by $v_d^{(E)}$, we have:
\begin{equation}
\langle L \rangle = v_d^{(E)} \left( \pi^{(E)} + \frac{v_d^{(C)}}{v_d^{(E)}} \left( \pi^{(S)} + \pi^{(P)} \right)  + \frac{[S(x)] + \mathcal{M}_E [P(x)]}{a_E + b_E [S(x)] + c_E [P(x)]} \right)
\end{equation}
where $a_E$, $b_E$, $c_E$ and $\mathcal{M}_E$ are the same quantities defined in Eqs.~\eqref{p1}, \eqref{p2} and \eqref{p3}, where $v_d$ appearing in the definition of $\lambda_0$ has to be replaced with $v_d^{(E)}$. Hence, reconstructing the same denominator, and writing $[P(x)]$ in terms of the entropy change, we have:
\begin{equation}
\langle L \rangle = v_d^{(E)} \left( \frac{a_E + (b_E + c_E e^{-\Delta S_m} (\mathcal{R}^{\rm eq})^{-1}) \left( v_d^{(C)}/v_d^{(E)} \right) }{a_E + (b_E + c_E e^{-\Delta S_m} (\mathcal{R}^{\rm eq})^{-1}) [S(x)]} + \frac{[S(x)] (1 + \mathcal{M}_E e^{-\Delta S_m} (\mathcal{R}^{\rm eq})^{-1})}{a_E + (b_E + c_E e^{-\Delta S_m} (\mathcal{R}^{\rm eq})^{-1}) [S(x)]} \right)
\end{equation}
Adding and substracting $a_E v_d^{(C)}/v_d^{(E)}$, we obtain:
\begin{equation}
\langle L \rangle = v_d^{(E)} \left( \frac{a_E \left( 1 - v_d^{(C)}/v_d^{(E)} \right)}{a_E + (b_E + c_E e^{-\Delta S_m} (\mathcal{R}^{\rm eq})^{-1}) [S(x)]} + \frac{v_d^{(C)}}{v_d^{(E)}} + \frac{[S(x)] (1 + \mathcal{M}_E e^{-\Delta S_m} (\mathcal{R}^{\rm eq})^{-1})}{a_E + (b_E + c_E e^{-\Delta S_m} (\mathcal{R}^{\rm eq})^{-1}) [S(x)]} \right)
\end{equation}
In our setting, the substrate is a monotonically decreasing function of $x$. Chemotaxis appears when $\partial_x P^{ss}(x) < 0$, i.e. the stationary distribution exhibits the same monotonicity as the substrate concentration. Recalling that $P^{ss}(x) \propto \langle L(x) \rangle^{-1}$, this condition is met when:
\begin{equation}
1 > \frac{v_d^{(C)}}{v_d^{(E)}} + \frac{1 + \mathcal{M}_E e^{-\Delta S_m} (\mathcal{R}^{\rm eq})^{-1}}{b_E + c_E e^{-\Delta S_m} (\mathcal{R}^{\rm eq})^{-1}}
\end{equation}
This corresponds to Eq.~(11) of the main text. Notice that, in order to obtain Fig.~3 (see main text), we evaluate $\mathcal{M}_E$, $a_E$, $b_E$ and $c_E$ using values obtained from the fitting procedure. We remark that now the estimated parameter $\alpha_S = \lambda/\lambda_0$ (Eq.~(9) of the main text) has to be interpreted as a version of the Damk$\ddot{\rm o}$hler number in the presence of kicks, depending only on enzyme diffusion, since only $v_d^{(E)}$ appears in $\lambda_0$.

In the experimental setting previously discussed, the enzyme is much bigger than the substrate, thus all diffusion rates are the same. As a consequence, stationary chemotactic profiles of the enzyme cannot appear in this case.

\end{widetext}

\end{document}